\documentclass[journal]{IEEEtran}
 
\usepackage{graphicx}
\usepackage{booktabs}
\usepackage{subfigure}
\usepackage{overpic}
\usepackage{amsmath}
\usepackage{flushend}
\usepackage{amssymb}
\usepackage{multirow}
\usepackage{makecell}
\usepackage{color, soul}
\soulregister\cite7

\usepackage[ruled]{algorithm2e}
\usepackage{hyperref}
\hypersetup{
	colorlinks=true,
	linkcolor=blue,
	urlcolor=blue,
	citecolor=blue}

\usepackage{colortbl}
\usepackage[dvipsnames]{xcolor}
\definecolor{Bg}{HTML}{e0f1ff}

\begin{document}

\title{Hierarchical Attention and Parallel Filter Fusion Network for Multi-Source Data Classification}
\author{Han Luo, Feng Gao, Junyu Dong, and Lin Qi
\thanks{This work was supported in part by the National Science and Technology Major Project under Grant 2022ZD0117201, and in part by the Natural Science Foundation of Qingdao under Grant 23-2-1-222-ZYYD-JCH. (\textit{Corresponding author: Feng Gao})

Han Luo, Feng Gao, Junyu Dong, Lin Qi are with the School of Information Science and Engineering, Ocean University of China, Qingdao 266100, China.}}

\markboth{IEEE GEOSCIENCE AND REMOTE SENSING LETTERS}%
{Shell}

\maketitle

\begin{abstract}

Hyperspectral image (HSI) and synthetic aperture radar (SAR) data joint classification is a crucial and yet challenging task in the field of remote sensing image interpretation. However, feature modeling in existing methods is deficient to exploit the abundant global, spectral, and local features simultaneously, leading to sub-optimal  classification performance. To solve the problem, we propose a hierarchical attention and parallel filter fusion network for multi-source data classification. Concretely, we design a hierarchical attention module for hyperspectral feature extraction. This module integrates global, spectral, and local features simultaneously to provide more comprehensive feature representation. In addition, we develop parallel filter fusion module which enhances cross-modal feature interactions among different spatial locations in the frequency domain. Extensive experiments on two multi-source remote sensing data classification datasets verify the superiority of our proposed method over current state-of-the-art classification approaches. Specifically, our proposed method achieves 91.44\% and 80.51\% of overall accuracy (OA) on the respective datasets, highlighting its superior performance.

\end{abstract}


\IEEEpeerreviewmaketitle

\section{Introduction}

\IEEEPARstart{W}{ith} the advancement of earth observation technology and satellite sensor platforms, a large amount of remote sensing data has been obtained. Among these data, Hyperspectral images (HSIs) have received a lot of attention due to their rich spectral information and have been widely used for land cover classification \cite{luo24tgrs}. However, spectral mixing occurs when HSI contains multiple land cover types, and it affects the HSI classification performance.

To solve the problem,  synthetic aperture radar (SAR) image is often used to  provide complementary information for HSI, since the SAR sensor is particularly useful in cloudy or hazy environments when HSI sensors may be limited. By combining HSI with SAR data, land cover classification methods can mitigate the effects of atmospheric interference, leading to more robust classification results. Therefore, in this letter, we mainly focus on HSI and SAR data joint classification.

Traditionally, attribute and extinction profile \cite{ep17jstars} are used for multisource data feature extraction. Xia et al. \cite{xia20grsl} presented graph fusion-based method, in which morphological filters were used to the key components of cross-modal data. Then, spectral, spatial, and texture features were projected to a lower subspace to compute joint features. However, traditional feature extraction methods are limited in extracting high-level semantic information from raw data.

Recently, many deep learning-based HSI and SAR data joint classification methods have been proposed. Wang et al. \cite{wjj23tnnls} presented a multi-scale interactive fusion network for multi-source data classification. Multi-scale features are extracted and fused via global dependence fusion module. Li et al. \cite{lw23tnnls} presented an asymmetric feature fusion network, in which a feature calibration module is designed to exploit the spatial dependence of multisource features. In addition, graph neural network \cite{yangjy24tgrs}, bilinear fusion \cite{songx23tgrs}, and adversarial learning \cite{advlearn23tgrs} are employed for multi-source remote sensing data classification.

\begin{figure*}[ht]
\centering
\includegraphics [width=7.0in]{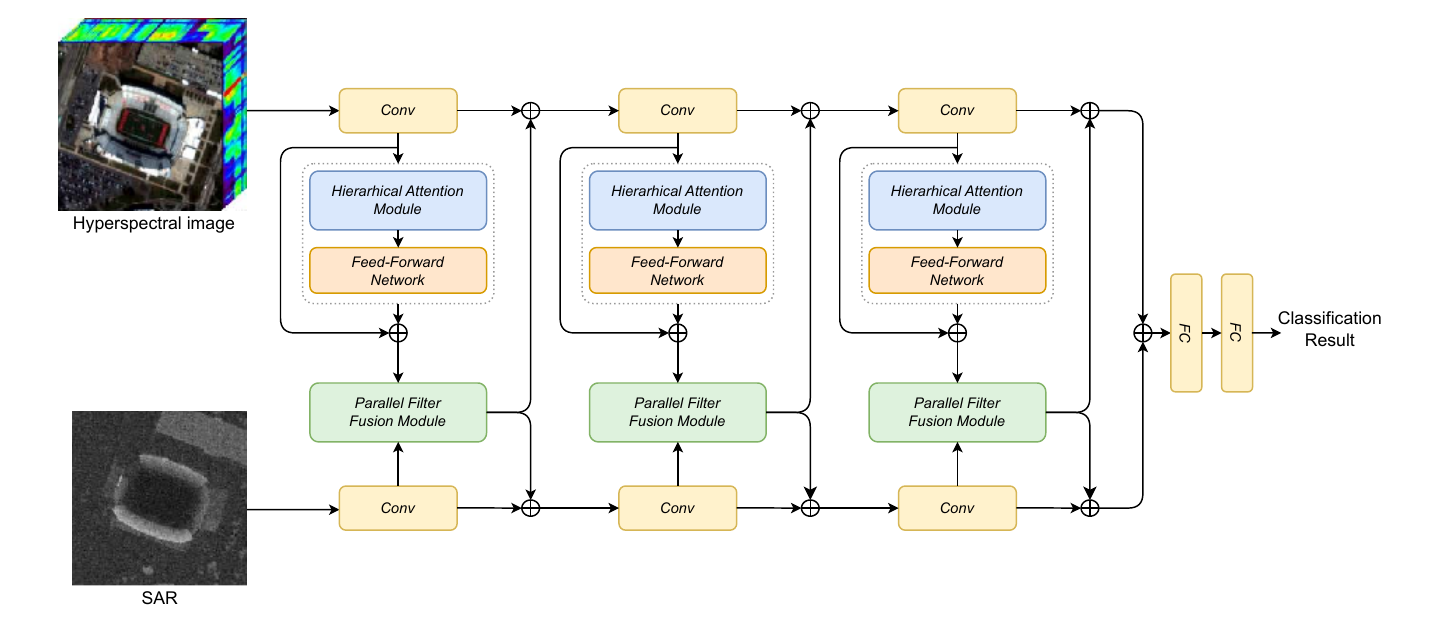}
\caption{Overview of the proposed HAPNet. In the HAPNet Block, multi-level spatial signals are fused through Hierarchical Attention Module (HAM), and the ability to  interact and fuse feature information is enhanced through the Parallel Filter Fusion Module (PFFM).} 
\label{fig_network}
\end{figure*}

Although existing methods have achieved remarkable performance, it is non-trivial to build a robust classification model due to the following two challenges: \textbf{1) Multi-granularity feature modeling for HSI data.} Cross-scale feature extraction is important for HSI feature representation. Existing methods can hardly model information at multiple granularities. Hence, how to modeling global, spectral, and local features simultaneously for HSI is a non-trivial task. \textbf{2) How to enhance cross-modal feature interactions between HSI and SAR data.} Existing multi-source image classification methods commonly capture multi-source feature interactions in the spatial domain. Feature interactions in the frequency domain is rarely explored. How to uncover the cross-modal feature interactions in the frequency domain is of great importance.

To overcome the above limitations, we propose a \underline{\textbf{H}}ierarchical \underline{\textbf{A}}ttention and \underline{\textbf{P}}arallel filter fusion \underline{\textbf{Net}}work, dubbed as HAPNet. Specifically, to efficiently extract multi-scale information, we design a Hierarchical Attention Module (HAM) to capture global, spectral and local information from HSI simultaneously. In addition, to enhance cross-modal feature interactions between HSI and SAR data, we propose a Parallel Filter Fusion Module (PFFM) for feature fusion. The feature interactions between HSI and SAR data are modeled as a set of learnable global filters which are applied to the spectrum of the input features. Therefore, cross-modal feature interactions among spatial locations are enhanced in the frequency domain. Extensive experiments on two HSI and SAR datasets have fully validated that our proposed method is superior to other state-of-the-art competitors.

Our main contributions can be summarized as follows:

\begin{itemize}

\item We present the HAM as an enhancement to the self-attention mechanism. This module integrates global, spectral, and local features to provide more comprehensive feature representation for multi-source remote sensing image classification.

\item We develop PFFM to enhance cross-modal feature interactions in the frequency domain. The module enhances feature interactions among different spatial locations in the frequency domain, and thus effectively improves the classification performance.

\end{itemize}

\section{Methodology}

As illustrated in Fig. \ref{fig_network}, the proposed HAPNet comprises HSI feature encoder, SAR feature encoder, feature fusion modules, and a classifier. Three hierarchical attention blocks are employed for HSI feature encoder, and three convolutional layers are used for SAR feature encoder. Subsequently, features from each level, derived from two encoder branches, are processed through feature fusion module. Finally, the fused features are fed into two fully-connected layers to generate the outcome. It should be noted that feature expansion is used here, and the expansion ratio is set to 2. In the HSI feature encoder, Principle Component Analysis (PCA) is employed to select the best 30 spectral bands. As illustrated in Fig. \ref{fig_network}, HAM and PFFM are the key components to improve the classification results. Subsequently, we provide in-depth descriptions of both modules.

\begin{figure}[ht]
\centering
\includegraphics [width=3.3in]{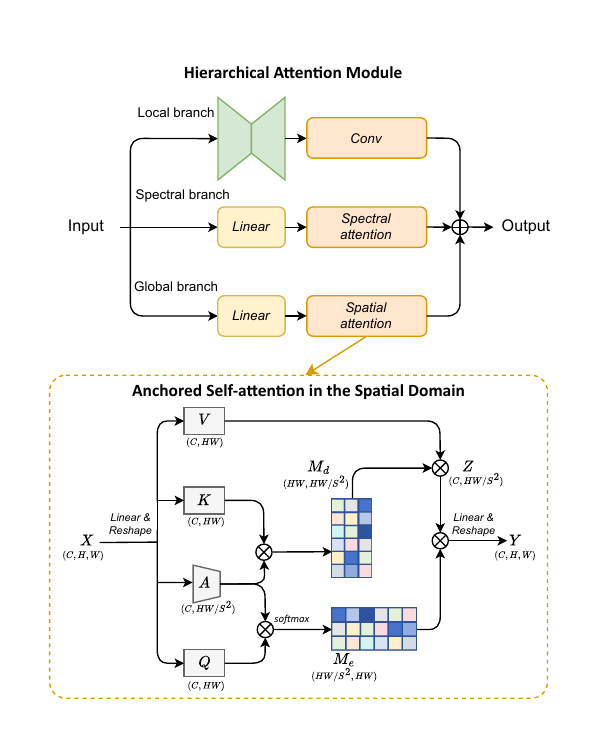}
\caption{Illustration of Hierarchical Attention Module (HAM). Global, spectral and local features are modeled in parallel. The global and spectral features are extracted via anchored self-attention in the spatial and spectral dimensions, respectively. The local features captures local structures via depth-wise convolutions.}
\label{fig_ham}
\end{figure}

\subsection{Hierarchical Attention Module (HAM)}

As depicted in Fig. \ref{fig_ham}, HAM is the key component that provides the hierarchical feature modeling capacity in the global, spectral, and local range. HAM first split the input feature into three branches. The global branch models long-range feature dependencies in the spatial dimension via anchored self-attention \cite{li23cvpr}. The spectral branch models long-range feature dependencies in the spectral dimension via shifted windows attention. The local branch captures local structures in the input feature via depth-wise convolutions. Features maps generated from the global, spectral, and local branch are combined via element-wise summation, and finally fed to the FFN to enhance the non-linear feature transformation. It should be noted that HAM is an extension of the self-attention mechanism. HAM provides a more comprehensive multi-granularity feature modeling approach through multi-scale and parallel feature processing. HAM captures global, spectral, and local information simultaneously, while the self-attention only captures global feature dependencies.

\textbf{Global and Spectral Feature Modeling.} In global and spectral branches, we use anchor self-attention for feature modeling. As depicted in Fig. \ref{fig_ham}, the triplet of $\mathbf{Q}$, $\mathbf{K}$, $\mathbf{V}$ is generated by linear projections. Then, an average pooling layer is implemented to reduce the spatial/spectral dimensions to generate $\textbf{A}$. The spatial/spectral dimension is down-scaled by a factor of $s$. 

To be specific, the anchor self-attention is computed as follows:

\begin{equation}
\mathbf{Y}=\mathbf{M_e}\cdot\mathbf{Z}=\mathbf{M_e}\cdot\left(\mathbf{M_d}\cdot\mathbf{V}\right),
\end{equation}
\begin{equation}
\mathbf{M_d}=\text{Softmax}\left(\mathbf{A}\cdot\mathbf{K}^T/\sqrt{d}\right),
\end{equation}
\begin{equation}
\mathbf{M_e}=\text{Softmax}\left(\mathbf{Q}\cdot\mathbf{A}^T/\sqrt{d}\right),
\end{equation}
where $\mathbf{A}$ is the anchor, $\mathbf{M}_e$ denotes the attention maps between the query-anchor pair, and $\mathbf{M}_d$ denotes the attention maps between the anchor-key pair. 

\textbf{Local Feature Modeling.} In local branch, channel-attention enhanced convolution is used for local feature modeling. Specifically, two depth-wise convolution layers are employed for feature extraction, and then channel-wise attention is used for feature enhancement.

\subsection{Parallel Filter Fusion Module (PFFM)}

\begin{figure}
  \centering
  \includegraphics[width=3.3in]{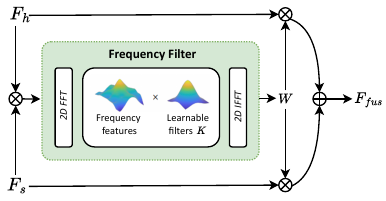}
  \caption{Illustration of the Parallel Filter Fusion Network (PFFM).}
  \label{PFFM}
\end{figure}

To enhance cross-model feature interactions, we propose the PFFM to adaptively fuse HSI and SAR features. It integrates Fourier transform-based filters into existing feature fusion network. As shown in Fig. \ref{PFFM}, the proposed PFFM comprises of three parallel paths. We use $F_h$ and $F_s$ to represent features from HSI and SAR, respectively. First, we use element-wise multiplication between $F_h$ and $F_s$ to generate $F_f$. Then, $F_f$ is fed into the global filter \cite{gfnet23tpami} to generate the attention weights $\mathbf{W}$. This process can be formulated as follows:
\begin{equation}
\mathbf{W}=\mathcal{F}^{-1} ( K\odot \mathcal{F} ( F_h \otimes F_s ) ),
\end{equation}
where $\mathcal{F}$ denotes the 2D FFT, $K$ is a learnable filter, $\mathcal{F}^{-1}$ denotes the inverse FFT. Finally, The enhanced features are combined via element-wise summation as follows:
\begin{equation}
    F_{fus}= (\mathbf{W} \otimes F_h) \oplus (\mathbf{W} \otimes F_s).
\end{equation}

The proposed PFFM use global filters to capture the feature interactions in the frequency domain. Therefore, cross-modal spectrum interactions are explored, which effectively improves the accuracy of multi-source data joint classification.

\begin{figure}[htbp]
  \centering
  \includegraphics[width=3.3in]{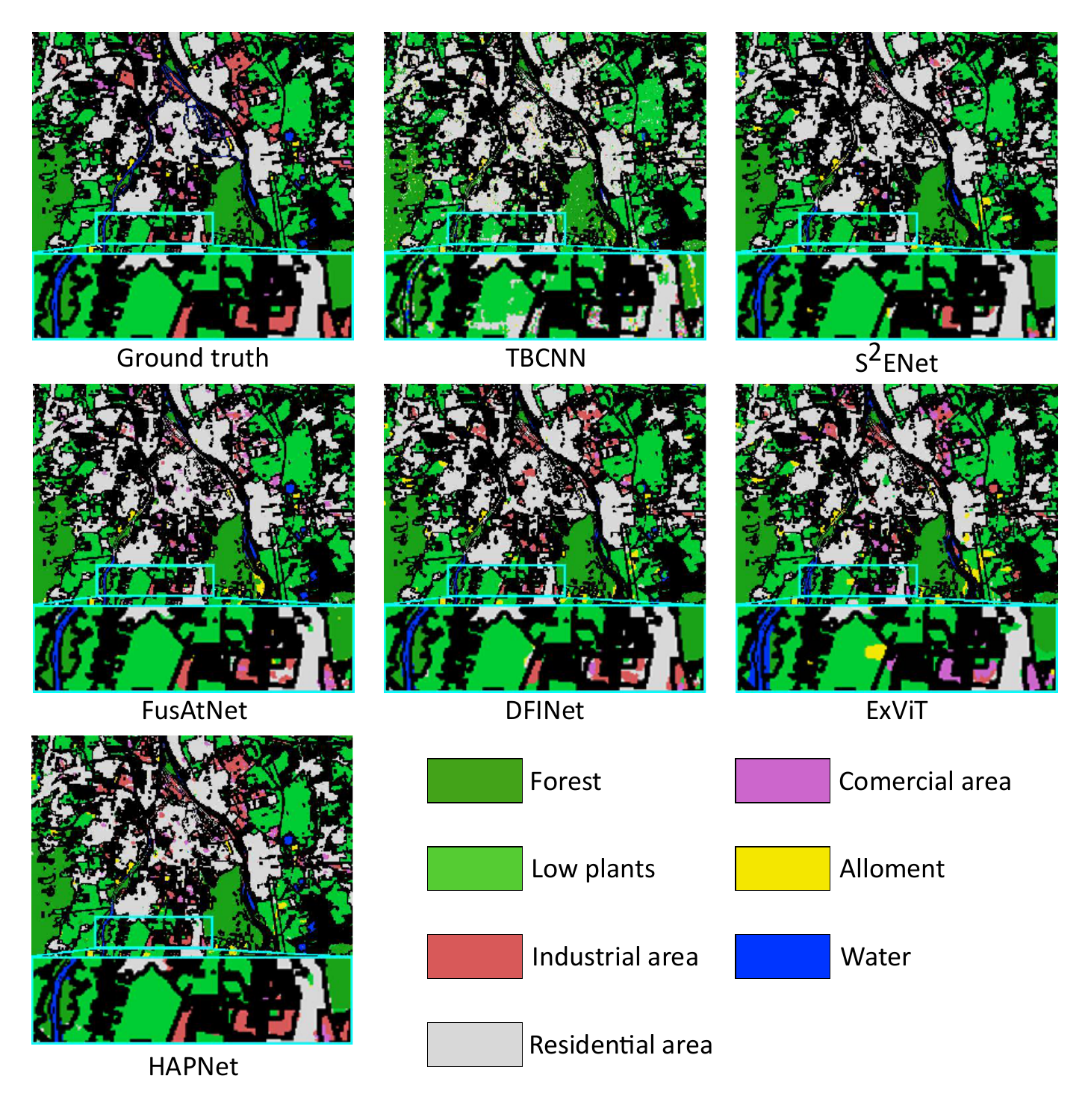}
  \caption{Classification results of different methods on the Augsburg dataset.}
  \label{fig_a}
\end{figure}

\begin{figure*}[htbp]
  \centering
  \includegraphics[width=6in]{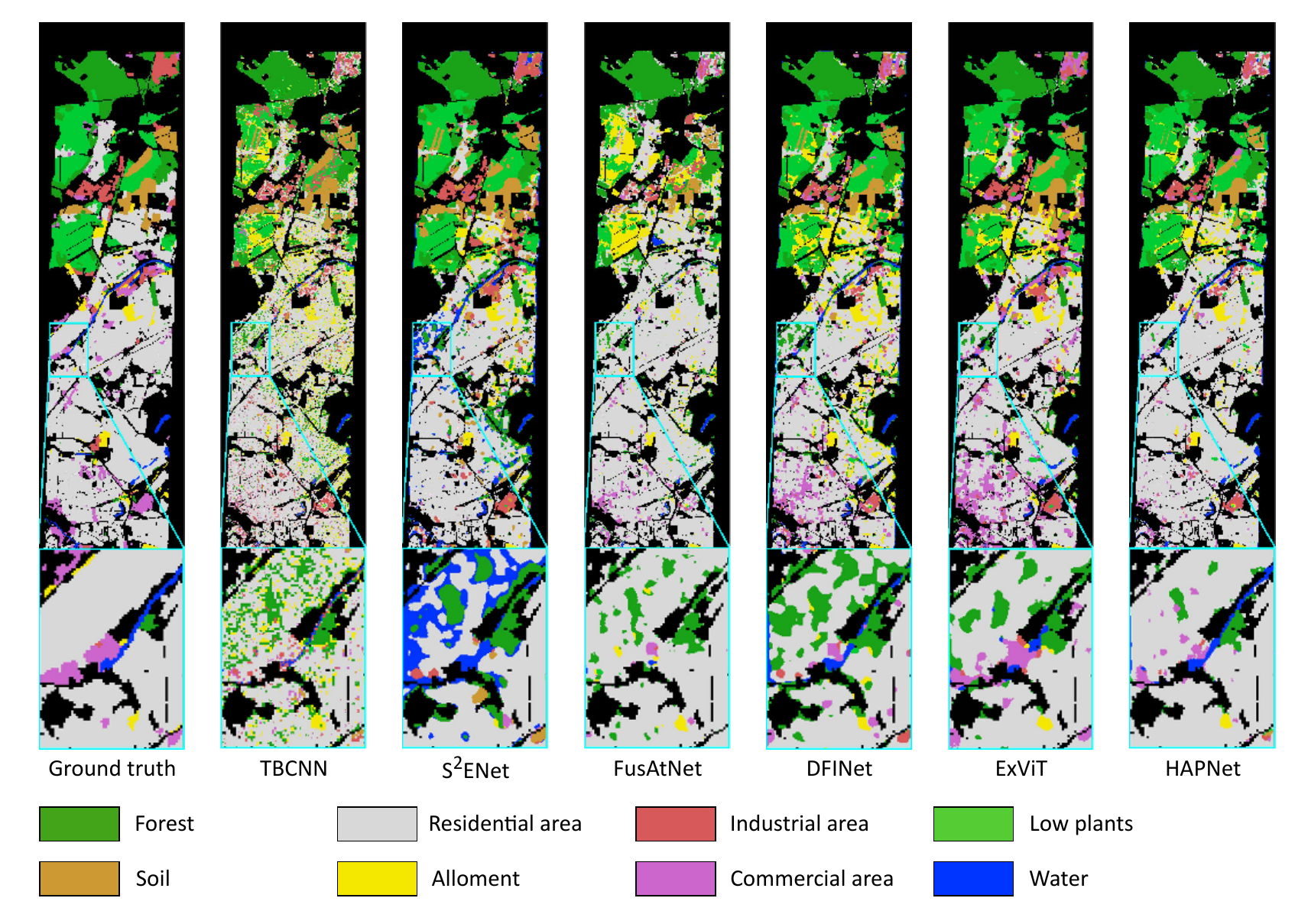}
  \caption{Classification results of different methods on the Berlin dataset.}
  \label{fig_b}
\end{figure*}

\section{Experimental Results and Analysis}

\subsection{Datasets and Experimental Setting}

We evaluate the performance of the proposed HAPNet on two HSI and SAR datasets. Specifically, the first dataset is the Augsburg dataset \cite{hong21isprs}. The HSI is captured by DAS-EOC, German Aerospace Center (DLR), over the city of Augsburg, Germany. The SAR data is collected by the Sentinel-1 sensor. For the HSI, there are 332 $\times$ 485 pixels and 180 spectral bands ranging from 0.4 to 2.5 $\mu$m. There are 7 distinct land cover classes in the ground truth. The second dataset is the Berlin dataset. The dataset contains 797 $\times$ 220 pixels, and the HSI contains 244 spectral bands. The are 8 distinct land cover classes in the ground truth. Due to the number of spectral bands in the Augsburg dataset and Berlin dataset being 180 and 244 respectively, to retain key feature information and reduce the input data volume of the neural network to improve its computational efficiency, we choose the PCA method to reduce the dimensionality of the spectral features in hyperspectral images to 30 channels. We use the standard training and test sets in both datasets. Specifically, 761 training samples are used on the Augsburg dataset, and 2820 training samples are used on the Berlin dataset.

To demonstrate the effectiveness of the proposed HAPNet, six state-of-the-art methods are selected for comparison: TBCNN \cite{tbcnn}, S$^2$ENet \cite{fs22grsl}, ExViT \cite{exvit}, FusAtNet \cite{fusatnet} and DFINet \cite{gyh22tgrs}. These methods are evaluated by visual comparison and quantitative metrics of individual class accuracy, overall accuracy (OA), average accuracy (AA), and Kappa coefficient. OA denotes the ratio of the total number of correctly classified pixels to the total number of pixels in the dataset, and it provides an overall assessment of the classification accuracy. AA provides a balanced measure by taking into account the accuracy for each class. Kappa provides a more robust measure of accuracy by taking into account the possibility of agreement by random chance.

The proposed HAPNet was conducted on NVIDIA RTX 3090 GPU. The training phase spanned over 100 epochs. The Adam optimizer is used with the learning rate of 0.0003. The batch size is set as 128. The input patch size for the proposed HAPNet is $11\times11$ pixels.

\subsection{Experimental Results and Discussion}

\begin{table}[ht]
\centering
\caption{Experimental results on the Augsburg dataset.}
\label{table_a}
\resizebox{0.95\linewidth}{!}{
\begin{tabular}{c|ccccccc}
\hline\toprule
Class  & TBCNN & S$^2$ENet & FusAtNet & DFINet & ExViT & HAPNet \\ 
\midrule
Forest           & 90.88 & \textbf{98.10} & 93.78 & 97.38 & 90.04 & 96.57 \\
Residential area & 93.89 & \textbf{99.08} & 97.58 & 98.37 & 95.44 & 95.78 \\
Industrail area  & 8.28  & 12.19 & 26.48 & 61.31 & 34.58 & \textbf{68.02} \\
Low plants       & 91.97 & 91.78 & \textbf{97.67} & 92.63 & 90.68 & 94.83 \\
Allotment        & 38.24 & 45.12 & 52.77 & 49.33 & 51.82 & \textbf{64.24} \\
Commercial area  & 1.40  & 1.22  & 24.66 & 3.54  & \textbf{28.63} & 18.08 \\
Water            & 10.82 & 24.09 & 47.51 & 26.61 & 17.65  & \textbf{48.00} \\
\midrule
OA               & 84.53 & 88.22 & 90.62 & 90.66 & 86.65  & \textbf{91.44} \\
AA               & 47.92 & 53.08 & 62.92 & 61.30 & 58.40  & \textbf{69.36} \\
Kappa            & 77.13 & 82.61 & 86.33 & 86.47 & 80.79  & \textbf{87.75} \\ 
\bottomrule\hline
\end{tabular}}
\end{table}

\begin{table}[ht]
\centering
\caption{Experimental results on the Berlin dataset.}
\label{table_b}
\resizebox{0.95\linewidth}{!}{
\begin{tabular}{c|ccccccc}
\hline\toprule
Class  & TBCNN & S$^2$ENet & FusAtNet & DFINet & ExViT & HAPNet \\ 
\midrule
Forest & 81.75 & 81.09 & \textbf{86.24} & 80.90 & 78.01 & 82.84 \\
Residential area & 76.26 & 73.05 & \textbf{91.38} & 72.81 & 74.05 & 89.47 \\
Industrial area  & 39.67 & \textbf{62.61} & 19.76 & 38.89 & 39.48 & 47.43 \\
Low plant & 49.78 & 82.82 & 20.00 & 78.09 & \textbf{84.15} & 81.71 \\
Soil & \textbf{89.42} & 86.41 & 48.72 & 73.48 & 88.03 & 71.12 \\
Allotment & 54.36 & 54.61 & 38.89 & \textbf{72.54} & 70.00 & 67.02 \\
Commercial area  & 4.65  & 2.56  & 18.47 & 22.80 & \textbf{38.18} & 24.74 \\
Water & 41.93 & 75.96 & 29.61 & 68.15 & 56.41 & \textbf{77.19} \\
\midrule
OA & 67.60 & 71.08 & 70.91 & 70.33 & 72.63 &\textbf{80.51} \\
AA & 54.72 & 64.88 & 44.13 & 63.45 & 66.04 & \textbf{66.44} \\
Kappa & 50.96 & 58.02 & 51.07 & 56.90 & 60.48 & \textbf{68.77}\\ \bottomrule\hline
\end{tabular}}
\end{table}

\textit{Results on the Augsburg dataset.} The classification maps of different methods on the Augsburg dataset is shown in Fig. \ref{fig_a}, and the corresponding quantitative evaluations are illustrated in Table \ref{table_a}. It should be noted that all the methods for comparison are designed for multi-source image joint classification. As can be seen in Table \ref{table_a}, the OA value for the proposed HAPNet achieves 91.44\%. It surpasses existing methods by at least 1.04\%. This indicates that HAPNet effectively integrates global, spectral, and local features, providing a more comprehensive feature representation for multi-source data classification. Specifically, HAPNet performs better than the S$^2$ENet, which is a self-attention-based method. It is evident that hierarchical attention provides better feature modeling capabilities than self-attention mechanism. Additionally, HAPNet performs better than the DFINet in which cross-attention fusion is employed. It demonstrated that cross-modal feature fusion in the frequency domain is more efficient than cross-attention.

\textit{Results on the Berlin dataset.} The classification results on the Berlin dataset is shown in Fig. \ref{fig_b}, and the corresponding quantitative evaluations are illustrated in Table \ref{table_b}. The proposed HAPNet achieves 80.51$\%$, 66.44$\%$ and 68.77$\%$ for OA, AA and Kappa, respectively. We found that Berlin dataset is more challenging compared to the Ausburg dataset. Nevertheless, the proposed HAPNet achieves the best performance on `Water' class. The reason might be that PFFM complements the water-sensitive features in SAR data with the HSI features. At the same time, the classification results of the other classes also achieve satisfying results. As a result, the proposed HAPNet performs better than the other methods on the Berlin dataset.

\textit{Computational Cost Analysis.} The computational costs of different methods are shown in Table \ref{table_comp}. As can be observed that, only ExViT is more computational efficient than the proposed HAPNet. However, the classification results of ExViT is not satisfying on both datasets. Compared with the other methods, the computational cost of the proposed HAPNet is quite competitive.

\begin{table}[ht]
\centering
\caption{Computational costs of different methods.}
\label{table_comp}
\resizebox{0.95\linewidth}{!}{
\begin{tabular}{c|ccccccc}
\hline\toprule
  & TBCNN & S$^2$ENet & FusAtNet & DFINet & ExViT & HAPNet \\ 
\midrule
FLOPs & 135.2M & 107.9M & 670.6M & 297.5M & 53.6M & 103.7M \\
\bottomrule\hline
\end{tabular}}
\end{table}

\subsection{Ablation Study}

We conduct a series of ablation experiments on two datasets to validate the effectiveness of the HAM and PFFM. We design two variants of HAPNet, i.e. without the HAM (w/o HAM) and without the PFFM (w/o PFFM). The experimental results are shown in Table \ref{table_ablation}. We find that the HAPNet always achieves better performance than its two variants on the two datasets. This demonstrates the necessity of the HAM and PFFM designed in HAPNet.

\begin{table}[ht]
\centering
\caption{Ablation study of the proposed HAPNet.}
\label{table_ablation}
\begin{tabular}{c|ccc} 
\toprule
\multirow{2}{*}{Method}
    & \multicolumn{2}{c}{OA on different datasets ($\%$)} \\ \cmidrule{2-3}
& Ausburg & Berlin\\ 
\midrule
HAPNet w/o HAM   & 90.35  & 74.49 \\  
HAPNet w/o PFFM  & 89.80  & 76.75 \\ 
Proposed HAPNet & \textbf{91.44}  & \textbf{80.51}  \\  
\bottomrule
\end{tabular}
\end{table}

\section{Conclusions}

In this letter, we developed an HSI and SAR data joint classification method which can model multi-granularity features simultaneously. To achieve this, we design HAM which integrates global, spectral, and local features to provide more comprehensive feature representation. In addition, we develop PFFM to enhance cross-modal feature interactions in the frequency domain. The module enhances feature interactions among different spatial locations in the frequency domain, and thus effectively improves the classification performance. Our HAPNet achieves very competitive performance with state-of-the-art methods on two HSI and SAR joint classification datasets.

\bibliographystyle{IEEEtran}
\bibliography{re} 

\end{document}